\documentclass[prl,aps,twocolumn,showpacs,nobibnotes,epsf]{revtex4}
\usepackage{graphicx}% Include figure files
\usepackage{dcolumn}% Align table columns on decimal point
\usepackage{bm}% bold math
\usepackage{SIunits}
\usepackage{booktabs}

\begin{document}
\title{ Superconductivity induced by oxygen deficiency in Sr-doped LaOFeAs }
\author{G. Wu, H. Chen, Y. L. Xie, Y. J. Yan, T. Wu, R. H. Liu, X. F. Wang, D. F. Fang, J. J. Ying}
\author{ X. H. Chen}
\altaffiliation{Corresponding author} \email{chenxh@ustc.edu.cn}
\affiliation{Hefei National Laboratory for Physical Science at
Microscale and Department of Physics, University of Science and
Technology of China, Hefei, Anhui 230026, P. R. China\\ }
\date{\today}

\begin{abstract}
We synthesized Sr-doped $La_{0.85}Sr_{0.15}OFeAs$ sample with single
phase, and systematically studied the effect of oxygen deficiency in
the Sr-doped LaOFeAs system. It is found that substitution of Sr for
La indeed induces the hole carrier evidenced by positive
thermoelectric power (TEP), but no bulk superconductivity is
observed. The superconductivity can be realized by annealing the
as-grown sample in vacuum to produce the oxygen deficiency. With
increasing the oxygen deficiency, the superconducting transition
temperature ($T_c$) increases and maximum $T_c$ reaches about 26 K
the same as that in La(O,F)FeAs. TEP dramatically changes from
positive to negative in the nonsuperconducting as-grown sample to
the superconducting samples with oxygen deficiency. While $R_H$ is
always negative for all samples (even for Sr-doped as grown sample).
It suggests that the $La_{0.85}Sr_{0.15}O_{1-\delta}FeAs$ is still
electron-type superconductor.
\end{abstract}


\vskip 300 pt

\maketitle

Since the discovery of high-transition temperature ($T_c$)
superconductivity in layered copper oxides, extensive efforts have
been devoted to explore the higher $T_c$ superconductivity. Layered
rare-earth metal oxypnictides LnOMPn (Ln=La, Pr, Ce, Sm; M=Fe, Co,
Ni, Ru and Pn=P and As) with ZrCuSiAs type
structure\cite{quebe,zimmer} have attracted great attention due to
the discovery of superconductivity at $T_c=26$ K in the iron-based
$LaO_{1-x}F_x$FeAs (x=0.05-0.12)\cite{yoichi}. $T_c$ was drastically
raised to more than 40 K beyond McMillan limitation of 39 K
predicted by BCS theory in $RO_{1-x}F_xFeAs$ by replacing La with
other trivalent R with smaller ionic radii \cite{chenxh,chen,ren}.
These discoveries have generated much interest for exploring novel
high temperature superconductor, and provided a new material base
for studying the origin of high temperature superconductivity.

Such high-$T_c$ iron pnictides adopts a layered structure of
alternating Fe-As and Ln-O layers with eight atoms in a tetragonal
unit cell. Similar to the cuprates, the Fe-As layer is thought to be
responsible for superconductivity, and Ln-O layer is carrier
reservoir layer to provide electron carrier. In order to induce the
electron carrier, three different ways have been used: (1)
substitution of flourine for oxygen \cite{yoichi,chenxh}; (2) to
produce oxygen deficiency\cite{ren1}; and (3) substitution of
$Th^{4+}$ for $Ln^{3+}$\cite{wang}. All these ways for inducing
electron carrier are limited in the carrier reservoir Ln-O layer by
substitution. The electron carrier induced transfers to Fe-As layer
to realize superconductivity. Superconductivity at 25 K has been
realized by hole doping with substituting $La^{3+}$ with $Sr^{2+}$
in LaOFeAs system\cite{wen}. The ternary iron arsenide $BaFe_2As_2$
shows superconductivity at 38 K by hole doping with partial
substitution of potassium for barium\cite{rotter}.  The undoped
material LaOFeAs shows an anomaly in resistivity at 150 K which is
associated with the structural transition or SDW
transition\cite{cruz,mcguire}. The SDW and the anomaly in
resistivity are suppressed, and superconductivity emerges with
increasing F doping\cite{dong,liu}. No anomaly in resistivity is
observed in optimal sample\cite{liu}. Therefore, the complete
suppression of the anomaly peak is an indication for inducing
carrier into system. Here we successfully prepared single phase
$La_{0.85}Sr_{0.15}OFeAs$, and systematically studied the effect of
oxygen deficiency on transport properties (resistivity, Hall
coefficient, and thermoelectric power). It is found that
substitution of $Sr^{2+}$ for $La^{3+}$ leads to the shift of the
anomaly peak to high temperature. Thermoelectric power changes sign
from negative to positive, while Hall coefficient keeps the same
sign and its magnitude deceases with Sr doping. The
superconductivity can be induced by annealing the as-grown sample in
vacuum to produce the oxygen deficiency. Both TEP and $R_H$ is
negative for the superconducting samples with oxygen deficiency. It
suggests that the $La_{0.85}Sr_{0.15}O_{1-\delta}FeAs$ is still
electron-type superconductor.

Polycrystalline samples with nominal composition LaOFeAs and
$La_{0.85}Sr_{0.15}O_{1-\delta}FeAs$ were synthesized by
conventional solid state reaction using high purity LaAs, SrCO$_3$,
Fe, As and $Fe_2O_3$ as starting materials. LaAs was obtained by
reacting La powder and As powder at 600 $^oC$ for 3 hours. The raw
materials were thoroughly grounded and pressed into pellets. The
pellets were wrapped into Ta foil and sealed in an evacuated quartz
tube. They are then annealed at 1160 $^oC$ for 40 hours. The sample
preparation process except for annealing was carried out in glove
box in which high pure argon atmosphere is filled. The
superconductivity is achieved with post-annealing of as-grown
samples $La_{0.85}Sr_{0.15}O_{1-\delta}FeAs$  for 2 hours and 4
hours in an high-evacuated quartz, respectively. Figure 1 shows the
XRD patterns for the polycrystalline samples LaOFeAs and
$La_{0.85}Sr_{0.15}O_{1-\delta}FeAs$ with different annealing time.
All peaks in XRD patterns can be well indexed to the tetragonal
ZrCuSiAs-type structure with single phase. The XRD patterns indicate
that all samples are stable for annealing in vacuum. Table I shows
lattice parameter for the sample $LaOFeAs$ and variation of a-axis
and c-axis lattice parameters with annealing time for the samples
$La_{0.85}Sr_{0.15}O_{1-\delta}FeAs$. It shows that both of a-axis
and c-axis lattice parameters decrease systematically with annealing
time. Increase in the annealing time in high vacuum leads to more
oxygen deficiency. It suggests that the lattice parameter a and c
decrease with increasing oxygen deficiency. Such variation of
lattice parameters with oxygen deficiency is consistent with the
results in $NdO_{1-\delta}FeAs$\cite{ren1}

\begin{figure}[t]
\includegraphics[width=9cm]{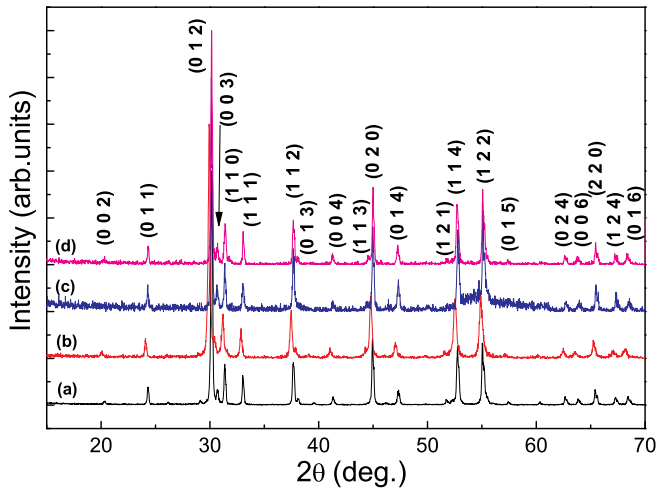}
\caption{(color online) X-ray diffraction patterns at room
temperature for the samples (a): $LaOFeAs$; (b): as-grown
$La_{0.85}Sr_{0.15}O_{1-\delta}FeAs$; (c): post-annealed
$La_{0.85}Sr_{0.15}O_{1-\delta}FeAs$ in high vacuum for 2 hours;
(d): post-annealed $La_{0.85}Sr_{0.15}O_{1-\delta}FeAs$ in high
vacuum for 4 hours.
\\}
\end{figure}

\begin{table}[h]
\caption{Lattice parameters for $LaOFeAs$ and
La$_{0.85}$Sr$_{0.15}$O$_{1-\delta}$FeAs samples with different
annealing time} \label{Table I} \centering
\begin{tabular}{ccccc}\\\hline
sample    &      a  ( {\AA} )     &     c  ( {\AA} )
\\\hline
pure  &    4.030(3) & 8.736(5) \\\hline
 annealing time  &      a  ( {\AA} )     &     c  ( {\AA}
)
\\\hline
   0 hrs              & 4.031(3)     & 8.749(5)     \\\hline
   2 hrs              & 4.027(3)     & 8.730(5)     \\\hline
   4 hrs              & 4.019(3)     & 8.723(5)     \\\hline
\end{tabular}
\end{table}

Temperature dependent resistivity for the samples LaOFeAs and
$La_{0.85}Sr_{0.15}O_{1-\delta}FeAs$ are shown in Fig.2. The undoped
compound LaOFeAs shows the same behavior as previous
report\cite{yoichi} and an anomaly at 150 K, which is believed to be
associated with the structural transition or SDW
transition.\cite{cruz,mcguire} As-grown Sr-doped LaOFeAs sample
shows different temperature dependent behavior from that observed in
the undoped LaOFeAs sample. The resistivity shows a linear
temperature dependence above a characteristic temperature of $\sim
165$ K, and steeply decreases with decreasing temperature below 165
K. Compared to the undoped LaOFeAs sample, the anomaly in
resistivity shifts to high temperature of 165 K associated with the
structural transition or SDW transition. The room-temperature
resistivity is about 13.7 m$\Omega$cm, being larger than that of
undoped LaOFeAs sample($\sim$ 5 m$\Omega$cm). However, a trace of
superconducting transition at $\sim 6$ K is observed as shown in
Fig.2. The resistivity shows a weak temperature dependent behavior
\begin{figure}[t]
\includegraphics[width=9cm]{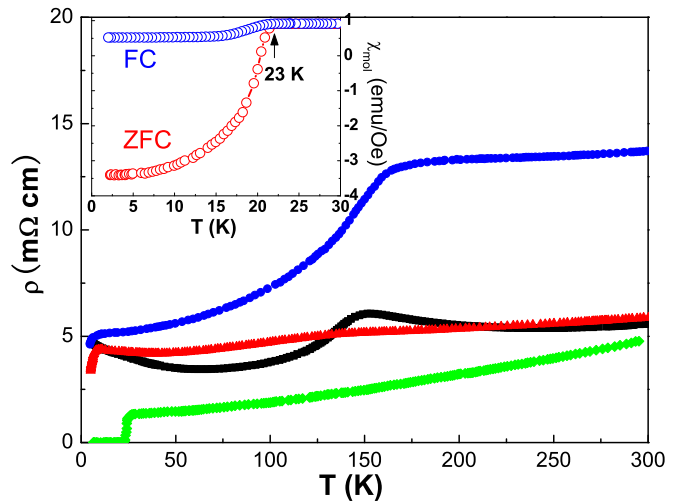}
\caption{(color online) Temperature dependence of resistivity for
the samples  LaOFeAs (squares); as-grown
$La_{0.85}Sr_{0.15}O_{1-\delta}FeAs$ (circles); post-annealed
$La_{0.85}Sr_{0.15}O_{1-\delta}FeAs$ in high vacuum for 2 hours (
triangles); post-annealed $La_{0.85}Sr_{0.15}O_{1-\delta}FeAs$ in
high vacuum for 4 hours (diamonds). The inset shows temperature
dependent susceptibility for $La_{0.85}Sr_{0.15}O_{1-\delta}FeAs$
annealed in high vacuum for 4 hours.
\\}
\end{figure}
for the sample obtained by annealing the as-grown
$La_{0.85}Sr_{0.15}O_{1-\delta}FeAs$ sample in a high vacuum for 2
hours.  The anomaly associated with the structural transition or SDW
transition is still observed at $\sim$ 140 K. Below 8 K, a
superconducting behavior is observed, and no zero resistivity is
obtained down to 4.2 K. The post-annealed
$La_{0.85}Sr_{0.15}O_{1-\delta}FeAs$ sample in high vacuum for 4
hours shows a well metallic behavior in resistivity and a sharp
superconducting transition occurs at 26 K and reaches to zero at
$\sim$ 23 K. The resistivity behavior is very similar to that of
$LaO_{0.89}F_{0.11}FeAs$\cite{yoichi}. The inset of Fig.2 shows the
temperature dependent susceptibility in zero field cooling (ZFC) and
field cooling (FC) for $La_{0.85}Sr_{0.15}O_{1-\delta}FeAs$ annealed
in high vacuum for 4 hours. Since the sample density is considerable
smaller than the theoretical value, we use 100.7 cm$^{3}$/mol and
get a superconducting fraction of $\approx$ 54$\%$ shielding. It
indicates a bulk superconductivity for
$La_{0.85}Sr_{0.15}O_{1-\delta}FeAs$ annealed  in high vacuum  for 4
hours.

The bulk superconductivity was realized by inducing oxygen
deficiency, the electron-doping is expected by introduction of
oxygen deficiency in Sr-doped $LaOFeAs$. In order to confirm this
expectation and provide the direct evidence, the thermoelectric
power and Hall coefficient are systematically measured. Temperature
dependent Hall coefficients for all samples are shown in Fig.3. The
sign of Hall coefficient for all samples are negative, indicating
electron-type carrier in these samples. $R_H$ shows a sharp increase
at the temperature of $\sim 150$ K associated with structure
transition or SDW transition for the undoped LaOFeAs. Which is
consistent with previous report\cite{mcguire}.
\begin{figure}
\includegraphics[width=9cm]{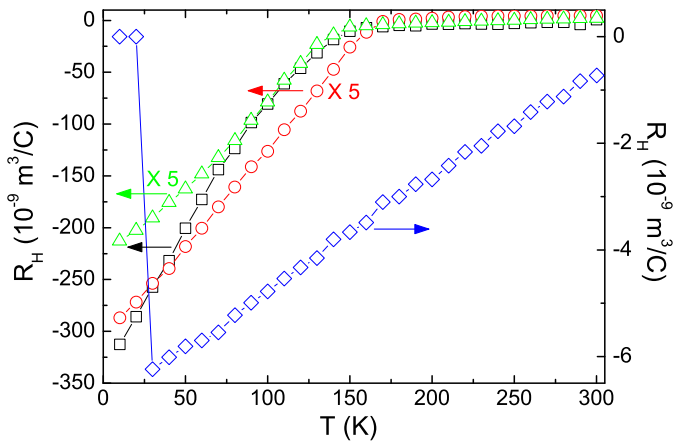}
\caption{(color online) Temperature dependence of Hall coefficient.
for the samples  LaOFeAs (squares); as-grown
$La_{0.85}Sr_{0.15}O_{1-\delta}FeAs$ (circles); post-annealed
$La_{0.85}Sr_{0.15}O_{1-\delta}FeAs$ in high vacuum for 2 hours (
triangles); post-annealed $La_{0.85}Sr_{0.15}O_{1-\delta}FeAs$ in
high vacuum for 4 hours (diamonds).
\\}
\end{figure}
$R_H$ of the sample Sr-doped $LaOFeAs$ shows similar temperature
dependence to that of pure $LaOFeAs$. But Sr-doping leads to an
decrease in magnitude of $R_H$, and seems to induce carrier into
system. Fig.3 clearly shows that the sharp increase in $R_H$ occurs
at $\sim 165$ K which coincides with the anomaly in resistivity
shown in Fig.2. This further indicates that the sharp increase in
$R_H$ arises from the SDW transition or structural phase transition.
Annealing in vacuum leads to a decrease in $R_H$, and shift of the
temperature corresponding to the sharp increase in $R_H$ to low
temperature. It indicates that the annealing in high vacuum induces
hole carrier and suppresses the SDW ordering, similar to effect of
F-doping in $SmO_{1-x}F_xFeAs$\cite{liu}. The superconducting sample
obtained by annealing in high vacuum for 4 hours shows a very small
$R_H$, and no sharp change in $R_H$ is observed above
superconducting transition. It indicates complete supression of SDW
transition or structural transition due to introduction of more
electron-carrier into system.

\begin{figure}[t]
\includegraphics[width=9cm]{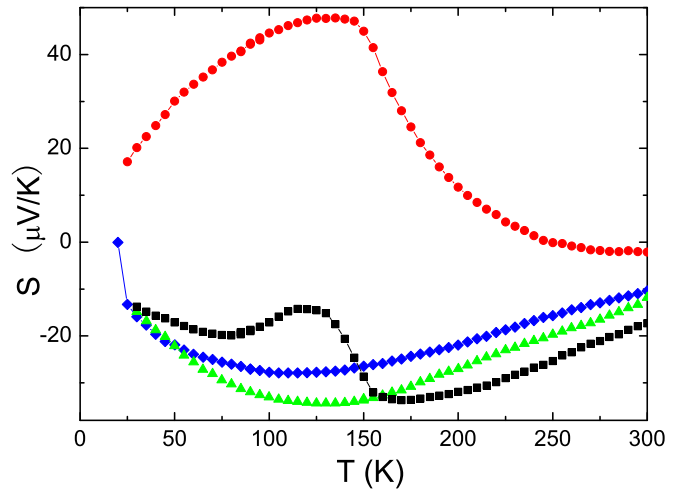}
\caption{(color online) Temperature dependent thermoelectric power
the samples  LaOFeAs (squares); as-grown
$La_{0.85}Sr_{0.15}O_{1-\delta}FeAs$ (circles); post-annealed
$La_{0.85}Sr_{0.15}O_{1-\delta}FeAs$ in high vacuum for 2 hours (
triangles); post-annealed $La_{0.85}Sr_{0.15}O_{1-\delta}FeAs$ in
high vacuum for 4 hours (diamonds)
\\}
\end{figure}

Figure 4 shows the temperature dependence of thermoelectric power
for all four samples. The parent compound $LaOFeAs$ shows similar
temperature dependent TEP to previous report\cite{mcguire}, below
the temperature ($\sim 150$ K) associated with SDW transition or
structural transition, a broad peak shows up. Sr-doping in $LaOFeAs$
leads to change the sign of TEP from negative to positive with
decreasing temperature at 250 K. Its temperature dependence shows a
typical behavior of a low carrier concentration materials which
contains electrons and holes as discussed in Ref.12. Positive TEP
indicates that the dominant carriers are holes, whereas $R_H$ has
shown that electrons dominate. In particular, $R_H$ is always
negative in entire temperature range. These experimental results are
consistent with those of the band structure calculations which
predicted that $LaOFeAs$ is a multi-band system. The opposite signs
of TEP ($>0$) and $R_H$ ($<0$) can be understood by considering that
the methods of averaging contributions of mutli-bands are different
for TEP and $R_H$. It is striking that the annealing in high vacuum
leads to change of TEP sign from positive to negative in entire
temperature range. The temperature dependence of TEP for the samples
obtained by annealing in vacuum is similar to that of
superconducting $LaO_{1-x}F_xFeAs$\cite{zhu}. At superconducting
transition temperature, TEP sharply drops to zero. It is intriguing
that the profile for temperature dependent TEP is similar to that of
low carrier concentration metals like undoped high-$T_c$ cuprates
except for negative sign. These results indicate that the dominate
carriers are electron in superconducting Sr-doped $LaOFeAs$.

It should be pointed that no superconductivity can be realized in
pure $LaO_{1-\delta}FeAs$ by annealing in high vacuum, this is
different from the report by Ren et al.\cite{ren1}. Ren et al.
reported that superconductivity can be obtained by high pressure
preparation in $LnO_{1-\delta}FeAs$. High pressure preparation could
produce enough oxygen deficiency, consequently induce the more
carrier concentration to realize the superconductivity. While the
annealing in vacuum cannot produce enough oxygen deficiency to
obtain superconductivity. Much more oxygen deficiency could lead to
the sample metastable. Such metastable sample can be obtained under
high pressure, while it cannot reach under high vacuum. It could be
reason why the superconductivity in $LnO_{1-\delta}FeAs$ can only be
obtain under pressure preparation. Sr-doping could play an important
role in removing oxygen from lattice to produce more oxygen
deficiency. This could reason why the superconductivity can be
realized in Sr-doped $La_{0.85}Sr_{0.15}O_{1-\delta}FeAs$ sample.
Both TEP and $R_H$ definitely indicate that the dominant carrier is
electron in superconducting $La_{0.85}Sr_{0.15}O_{1-\delta}FeAs$
system. So far, it seems that n-type carrier can be induced into the
system $LnOFeAs$ with single FeAs layer, while p-type carrier is
induced to the superconductors $Ba_{1-x}K_xFe_2As_2$ with double
FeAs layers\cite{wu}. It is different from the case of high-$T_c$
cuprates, it should be interesting issue.

\vspace*{2mm} {\bf Acknowledgment:} This work is supported by the
Nature Science Foundation of China and by the Ministry of Science
and Technology of China (973 project No: 2006CB601001) and by
National Basic Research Program of China (2006CB922005).

\

\end{document}